\documentclass[aps]{revtex4}

\usepackage{graphicx}
\usepackage{epsfig}
\usepackage[dvips]{color}
\usepackage{amssymb}

\newcommand{\be}{\begin{equation}}
\newcommand{\ee}{\end{equation}}
\newcommand{\ba}{\begin{eqnarray}}
\newcommand{\ea}{\end{eqnarray}}
\newcommand{\ban}{\begin{eqnarray*}}
\newcommand{\ean}{\end{eqnarray*}}

\newcommand{\ZZ}{{\mathbb Z}}

\newcommand{\braj}{{\langle {\bf j}\vert}}

\newcommand{\bk}{{\bf k}}
\newcommand{\bj}{{\bf j}}
\newcommand{\bi}{-\left[\frac d2\right]}
\newcommand{\bs}{\left[\frac {d-1}2\right]}

\newcommand{\one}{\leavevmode\hbox{\small1\normalsize\kern-.33em1}}

\begin{document}

\title{Uncertainty Relation for the Discrete Fourier Transform}

\author{Serge Massar}
\affiliation{Laboratoire d'Information Quantique, {C.P.} 225, Universit\'{e} Libre de Bruxelles (U.L.B.), Boulevard du Triomphe, 1050 Bruxelles, Belgium}

\author{Philippe Spindel}
\affiliation{Service de M\'ecanique et Gravitation, Universit\'e de Mons-Hainaut, 
Acad\'emie universitaire Wallonie-Bruxelles, 
Place duParc 20, BE-7000 Mons, Belgium 
}

\date{\today}

\begin{abstract}
We derive an uncertainty relation for two unitary operators which obey a commutation relation of the form $UV=e^{i\phi}VU$. Its most important application is to  constrain how much a quantum state can be localised simultaneously in two mutually unbiased bases related by a Discrete Fourier Transform. It provides an uncertainty relation which smoothly interpolates between the well known cases of the Pauli operators in 2 dimensions and the continuous variables position and momentum. This work also provides an uncertainty relation for modular variables, and could find applications in signal processing. In the finite dimensional case the minimum uncertainty states, discrete analogues of coherent and squeezed states, are minimum energy solutions of Harper's equation, a discrete version of the Harmonic oscillator equation.
\end{abstract}

\maketitle


\section{Introduction}

Uncertainty relations provide some of our most fundamental insights into quantum mechanics. They express the fact that non commuting observables cannot simultaneously have well defined values. This concept has no classical analogue, and thereofore underlies much of the conceptual differences between classical and quantum mechanics. For these reasons uncertainty relations have attracted a huge amount of attention.

The uncertainty principle was first understood by Heisenberg\cite{Heisenberg}, and formulated precisely by Kennard as\cite{Kennard}
\begin{equation}
\Delta x \Delta p \geq \frac{1}{2} \ .
\label{DxDp}
\end{equation}
Here $x$ and $p$ are the position and momentum observables, the variance of observable $A$ in state $|\psi\rangle$ is
\begin{equation}
\Delta A^2 = \langle \psi | A^2 | \psi \rangle - \langle \psi | A | \psi \rangle^2\ ,
\label{DA}
\end{equation}
and we work in units where $\hbar=1$, {\it i.e.} $[x,p]=i$. This relation was subsequently generalised by Robertson\cite{Robertson} to 
\begin{equation}
\Delta A \Delta B \geq \frac{1}{2} | \langle \psi | [A,B] | \psi \rangle|
\label{DADB}
\end{equation}
for any observables $A$ and $B$.

The relation eq. (\ref{DADB}) is however not always satisfactory. For instance uncertainty relations for  phase and number, or angle and angular momentum, are notoriously tricky, see
\cite{CarruthersNieto} for an excellent review. In the discrete case there has also been some important work.
First of all note that for spin $1/2$ particles, the uncertainty relations for the Pauli operators  (which cannot be deduced from eq. (\ref{DADB}), but can be easily be established from the definition $\Delta \sigma_x^2=1 - \langle \sigma_x \rangle^2$ and the constraint $\langle \sigma_x \rangle^2 + \langle \sigma_y \rangle^2 + \langle \sigma_z \rangle^2 \leq 1$ which is saturated for pure states) is 
\begin{equation}
\Delta \sigma_x^2 + \Delta \sigma_z^2   \geq 1 \ .
\label{DxDz}
\end{equation}
An important reinterpretation of eq. (\ref{DxDz}) is as
an uncertainty relation for Mach-Zehnder interferometers in which one relates the predictability of the path taken by the particle to the visibility of the interference fringes, see e.g. \cite{GY,JSV}. This has been extended to the case of multipath interferometers, see e.g. \cite{Durr,KKZE}.  
Finally we mention that other more information theoretic uncertainty relations, such as entropic uncertainty relations, have also been developped \cite{BBM,Deutsch,Kraus,MU}.

In the present work we derive uncertainty relations for two unitary operators that obey the commutation relation 
$U V = e^{i \phi} V U$. This uncertainty relation has several important applications: it provides an uncertainty relation for the Discrete Fourier Transform (DFT), and in this context provides a family of uncertainty relations that interpolate between the case of Pauli operators eq. (\ref{DxDz}) and that of position and momentum eq. (\ref{DxDp}); it also provides an uncertainty relation for modular variables; finally it should prove usefull in signal processing.

{ We also caracterise the quantum states with minimum uncertainty in two bases related by the DFT. These states are discrete analogues of the coherent and squeezed states that are so important in the study of continuous variable systems. They have already been studied previously\cite{O95,BCHKO}. They are minimum energy eigenstates of Harper's equation\cite{Harper}, a discrete version of the Harmonic Oscillator Hamiltonian for continuous variables.}

We begin by presenting the different applications, before stating and proving our results.

\section{Discrete Fourier Transform}

Mutually unbiased bases have been extensively studied because of their nice properties and potential applications in quantum information. For instance they can be useful for quantum key distribution \cite{BPP,CBKG}, for locking of quantum information\cite{DHLSMT}, for string committement\cite{BCHLW}. A particularly interesting case occurs when the two bases are related by a DFT:
$$
{|\widetilde\bk\rangle} = \sum_{j=\bi}^{\bs} 
\frac{e^{+i 2 \pi j k /d} } {\sqrt{d}}|\bj\rangle 
\; , \;
|\bj\rangle = \sum_{k=\bi}^{\bs} 
\frac{ e^{-i 2 \pi j k /d} } {\sqrt{d}}
{|\widetilde\bk\rangle}
$$
with $\langle \bj | \bj'\rangle = \delta_{j j'}$, ${\langle\widetilde \bk |}{\widetilde\bk'\rangle} = \delta_{k k'}$ and $j,j',k,k'=\bi,\ldots,\bs$.
This case find applications in the Pegg-Barnett approach to phase-number uncertainty relations\cite{PB}, and in multipath interferometers since the "symmetric multiport beam splitter" considered in \cite{KKZE} is just the DFT. 
The question we ask is: {\em How much can a state be simultaneously localised both in the $|\bj\rangle$ and the $ {|\widetilde\bk\rangle}$ bases?} 

Because of the cyclic invariance of the DFT, it is natural to use a measure of localisation which is invariant under cyclic permutations. To this end we introduce the unitary operators
\begin{eqnarray}
U = \sum_{j=\bi}^{\bs} e^{+i 2 \pi j  /d} |\bj\rangle \langle \bj | 
\ , \ 
V = \sum_{k=\bi}^{\bs} e^{-i 2 \pi k  /d}  {| \widetilde\bk\rangle} { \langle\widetilde \bk |} \label{V}
\end{eqnarray}
We shall measure the localisation in the two bases by the generalisation of eq. (\ref{DA}) to non hermitian operators:
\begin{eqnarray}
\Delta U^2 &=& \langle \psi | U^\dagger U | \psi \rangle - 
\langle \psi | U^\dagger | \psi \rangle\langle \psi | U | \psi \rangle =
1 - | \langle \psi | U | \psi \rangle |^2 \nonumber
\\
\Delta V^2 &=& 
1 - | \langle \psi | V | \psi \rangle |^2
\nonumber\\
\label{DU}
\end{eqnarray}
The uncertainties $\Delta U^2$ and $\Delta V^2$ are the discrete versions of the {\em dispersion} introduced in \cite{BP}, see also \cite{BFS}.
Note that we have
$0\leq \Delta U^2 \leq 1$ and $0\leq \Delta V^2 \leq 1$.

For further use let us collect here some important properties of the operators $U$ and $V$.
They can be written as
\begin{eqnarray}
U= \sum_{k=\bi}^{\bs}  {| \widetilde{\bf k +1} \rangle} {\langle\widetilde \bk |} 
\quad , \quad
V= \sum_{j=\bi}^{\bs}  |{\bf j+1} \rangle\langle \bj | 
\label{V2}
\end{eqnarray}
and obey the commutation relations
\begin{eqnarray}
U^nV^m &=& V^mU^n e^{+i2\pi  nm/d} \nonumber\\
 U^{\dagger n}V^m &=& V^mU^{\dagger n} e^{-i2\pi  nm/d}
\label{UVVU}
\end{eqnarray}
They also act as translation operators, since if
\begin{eqnarray}
|\psi\rangle \to U^a V^{-b} |\psi\rangle \ ,
\label{Trans}
\end{eqnarray}
then $\langle U \rangle \to e^{i 2 \pi b/d} \langle U \rangle$ and
$\langle V \rangle \to e^{i 2 \pi a/d} \langle V \rangle$.

Our motivation for developping an uncertainty relation for the $U$ and $V$ operators is that the DFT interpolates between two important limits. In the $d=2$ case we can identify $U=\sigma_x$ and $V=\sigma_z$ and the uncertainty relation eq. (\ref{DxDz}) applies.

And in the supplementary material we discuss in detail how  
in the limit $d\to \infty$ the DFT approximates the Continuous Fourier Transform (CFT).
The idea is to rewrite
$U=e^{i u \sqrt{2 \pi / d}}$ and $V=e^{i v \sqrt{2 \pi / d}}$, where $u$ and $v$ are Hermitian operators with eigenvalues $n \sqrt{2 \pi /d}$, $n\in\{\bi,\ldots,\bs\}$; and to consider states for which $1- \langle \psi |U |\psi \rangle=\mu$ and $1-\langle \psi |V |\psi \rangle=\mu'$ are both small complex numbers  ($|\mu|,|\mu'|\ll 1$).  This implies that $\Delta U^2=O(|\mu|)$ and $\Delta V^2=O(|\mu'|)$ are both very small. We then show that on such states one can 
approximate $U$ and $V$ by their  series expansions:
$U \simeq 1 + i \sqrt{\frac{2 \pi}{d}}u - \frac{ \pi}{d}u^2$
and
$V \simeq 1 + i \sqrt{\frac{2 \pi}{d}}v - \frac{ \pi}{d}v^2$.
This in turn implies that 
$\Delta U^2 \simeq \frac{2 \pi}{d}  (\langle u^2\rangle - \langle u \rangle^2 )$ and
$\Delta V^2 \simeq \frac{2 \pi}{d}  (\langle v^2\rangle - \langle v \rangle^2 )$, {\it i.e.} $\Delta U^2$ and $\Delta V^2$ are proportional to the uncertainty of the operators $u$ and $v$ in the sense of eq. (\ref{DA}).
Furthermore, inserting the joint expansion into eq.
(\ref{UVVU})  we obtain 
$uv - vu \simeq i$. Thus, when acting on this class of states,
$u$ and $v$ are analogues of the conjugate variables $x$ and $p$. 
It then follows from eq. (\ref{DxDp}) that 
$\Delta^2 U$ and $\Delta V^2$ cannot both be made arbitrarily small, since when the above conditions hold they must obey 
the constraint
\begin{equation}
\Delta U^2 \Delta V^2 
\geq \frac{\pi^2}{d^2} \ .
\label{DUDVwrong}
\end{equation}
Note however that eq. (\ref{DUDVwrong}) does not  hold when $\Delta U^2$ or $\Delta V^2$ are large. Indeed if we take states that are perfectly localised in one basis or in the other we have
\begin{eqnarray}
|\psi\rangle = |\bj\rangle &\Rightarrow& \Delta U =0 \quad \& \quad \Delta V=1 \label{U0V1}\\
|\psi\rangle = \tilde{| \bk\rangle} &\Rightarrow& \Delta U =1 \quad \& \quad \Delta V=0 \ . \label{U1V0}
\end{eqnarray}
One of our tasks is to find an uncertainty relation that correctly interpolates between the limits eq. (\ref{DUDVwrong}) and eqs. (\ref{U0V1}, \ref{U1V0}).

\section{Modular Variables}

An interesting generalisation of the commutation relation eq. (\ref{UVVU}) is provided by the 
translation operators 
$U = e^{i 2 \pi  x/L}$ and $ V = e^{-i 2 \pi p/P}$
which obey the commutation relations
 \begin{eqnarray}
U  V &=& VU  e^{i \Phi} \nonumber\\
U^\dagger  V &=& VU^\dagger  e^{-i \Phi}
\label{COMMod}
\end{eqnarray}
with $\Phi = 4 \pi^2 / LP$. In what follows we shall base our study on unitary operators that obey commutation relations of the type eq. (\ref{COMMod}), {\it i.e.} we allow arbitrary  values of $\Phi$.

The generators  $x \mbox{ mod} L$ and $p \mbox{ mod} P$ of the translation operators $U$ and $V$ are called modular variables. These were introduced in \cite{APP} as a tool for understanding non local phenomena in quantum mechanics. Our uncertainty relation for $U$ and $V$ thus also provides an uncertainty relation for the modular variables.

\section{Signal Processing}

Uncertainty relations for $U$ and $V$ operators also have implications for signal processing.

On the one hand discrete generalisation of the Q-function, the P-function and other discrete phase space functions always refer to a particular state. Minimum uncertainty states are thus natural candidates for these reference states, as discussed in detail in \cite{OBBD95,OWB96}. 

On the other hand we can express the quantum state $|\psi\rangle$ in the $|j\rangle$ basis
$
|\psi\rangle = \sum_j c_j |\bj\rangle
$
and re-interpret the $c_j$ as a discrete signal of period $d$ normalised to
$
\sum_j |c_\bj |^2 =1
$.
The Discrete Fourier Transform of the signal $c_j$ is
$
\tilde c_k = \frac{1}{\sqrt{d}}\sum_j e^{-i 2 \pi j k /d} c_j
$.

The fundamental theorem of signal processing, the Wiener-Kinchin theorem, states that the correlation function is the Fourier transform of the spectral intensity:
\begin{equation}
\sum_j c_{j+m}^* c_j = \sum_k e^{-i 2 \pi k m /d} |\tilde c_k |^2 = \langle \psi | V^m | \psi \rangle
\label{X}
\end{equation}
In the quantum language it corresponds to the two different expressions for $V$, eqs. (\ref{V}) and (\ref{V2}).

Similarly the expectation value of $U^n$ 
\begin{equation}
\sum_j |c_j |^2 e^{i 2 \pi j n /d} = \sum_k \tilde c_{k+n}^* \tilde c_k =\langle \psi | U^n | \psi \rangle
\label{Y}
\end{equation}
is the Fourier transform of the intensity time series.

In view of this correspondence, our main result stated below provides a constraint between the values of the correlation function (\ref{X}) and the Fourier Transform of the intensity time series (\ref{Y}). This kind of constraint should prove useful in signal processing, as it constrains what kinds of signals are possible, or what kind of wavelet bases one can construct.

\section{Results}

Our main result is:

{\em Theorem 1: Consider two unitary operators $U$ and $V$ which obey
\begin{equation}
UV = VU e^{i \Phi}
\ ,\ 
U^\dagger V = V U^\dagger e^{-i \Phi}
\ ,\  0\leq \Phi\leq \pi
\label{UVVU2THEO}\end{equation}
and define
\begin{eqnarray}
\Delta U^2 = 1 - |\langle \psi |U |\psi \rangle |^2 
\quad,\quad
\Delta V^2 = 1 - |\langle \psi |V |\psi \rangle |^2
\end{eqnarray}
which are trivially bounded by 
$ 0 \leq \Delta U^2 \leq 1$, $0 \leq \Delta V^2 \leq 1
$
and let
\begin{equation}
A=\tan \frac{\Phi}{2} \quad  , \quad 0 \leq A \leq + \infty \ .
\end{equation}
Then we have the bound
\begin{equation}
(1 + 2 A) \Delta U^2 \Delta V^2 
+ A^2 (\Delta U^2 + \Delta V^2) \geq A^2 \ .
\label{THEO}
\end{equation}
}
The proof of Theorem 1 is given in the Supplementary Material.

Let us note that Theorem 1 correctly yields the expected asymptotic behaviors. To study the $d\to \infty$ limit, rewrite eq.  (\ref{THEO}) as
$$  \frac {\Delta U^2 \Delta V^2}{A^2} \geq 1 - \left ( \Delta U^2 + \Delta V^2 + \frac{2}{A}\Delta U^2 \Delta V^2\right)
$$  
For large $d$ we have $A \simeq \Phi / 2 =  \pi /d \to 0$. 
We then recover eq. (\ref{DUDVwrong}) when the terms in parenthesis on the right hand side are negligible in front of $1$, that is when $\Delta U$ and $\Delta V$ are both sufficiently small.

 In addition eq. (\ref{THEO}) is saturated by the two particular points   eqs. (\ref{U0V1}) and (\ref{U1V0}). 

Finally eq. (\ref{THEO}) gives the correct behavior when $d=2$, eq. (\ref{DxDz}). Indeed $d=2$ is obtained as the limiting case $\Phi \to \pi$, corresponding to $A \to \infty$.
 
 However numerical investigations for small dimensionality
d show that, except for d = 2, the bound is
not tight, ie. there are no states which saturate eq. (\ref{THEO}),  see fig. \ref{fig:FourierGraph}.
On the other hand, as in \cite{O95}, a tight bound can be obtained implicitly as the minimum eigenvalue of a Hermitian Operator (Harper's equation), and the minimum uncertainty states are the associated eigenstates.
Too see this we change slightly our point of view, and 
instead of looking at the accessible region in the $\Delta U^2$, $\Delta V^2$ plane, we look at the accessible region in the $| \langle \psi | U | \psi \rangle |$, $| \langle \psi | V | \psi \rangle |$ plane. We state the following two results for finite dimensional spaces (leaving open the exact way in which they should be formulted for the infinite dimensional case):

 \begin{figure}[t]
  \includegraphics[scale=0.8]{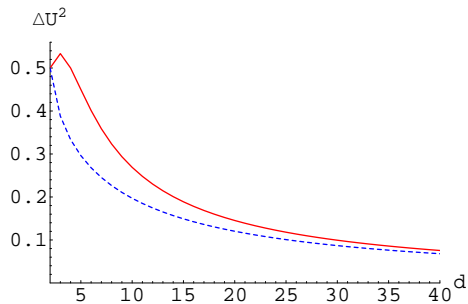}
     \caption{Minimum Uncertainty $\Delta U^2$ as a function of dimension $d$ when one imposes that $\Delta U^2=\Delta V^2$. The upper (red and continuous) curve  is the exact bound on $\Delta U^2$. It is obtained from the smallest eigenvalue of the operator eq. (\ref{Hamilt}) when $\theta=\pi/4$. Note that when $d=2$ and $d=4$ the exact bound is $\Delta U^2=1/2$ and that when $d=3$ the exact bound is larger than $1/2$, as noted in \cite{O95}. The lower (blue and dashed) curve is the bound obtained from the bound eq. (\ref{THEO}) upon imposing that $\Delta U^2=\Delta V^2$. The two curves coincide when $d=2$ and have the same asymptotic behavior $\Delta U^2 \geq \pi/d$ for large $d$.}
     \label{fig:FourierGraph}
     \end{figure}

{\em Theorem 2: Consider a $d$ dimensional Hilbert space, and two unitary operators $U$, $V$ acting on that space that obey the conditions of Theorem 1 with $\Phi = 2 \pi/d$. Then the maximum of
\begin{equation}
 \cos \theta |\langle \psi| U  |\psi \rangle| + \sin \theta |\langle \psi |  V |\psi  \rangle| , \quad 0\leq \theta \leq \pi/2
 \label{B1}
 \end{equation}
 is given by the smallest eigenvalue of the Hermitian operator
 \begin{equation}
 H= -\cos \theta C_U  - \sin \theta C_V  , \quad 0\leq \theta \leq \pi/2
 \label{Hamilt}
 \end{equation}
 where $C_U= (U + U^\dagger)/2$ and $C_V = (V+V^\dagger)/2$.
}

Note that Theorem 2 gives implicitly the boundary of the accessible region in the $|\langle U \rangle|$, $|\langle V \rangle|$ space (more precisely the convex hull of the accessible region). A comparison of the bound obtained from Theorem 1 and Theorem 2 in the case $\theta=\pi/4$ is given in Fig. \ref{fig:FourierGraph}.

A slight extension of the proof of Theorem 2 also provides a method to construct the states that saturate the uncertainty relation for $U$ and $V$:  

{\em Theorem 3: Consider a $d$ dimensional Hilbert space, two unitary operators $U$, $V$, and the Hermitian operator $H$, as described in the statements of Theorems 1 and 2. Denote by $h_{min}$ the smallest eigenvalue of $H$. Denote by $|\psi_{min}\rangle$ the eigenvector corresponding to the smallest eigenvalue of $H$. Then the unique states states that maximise 
 eq. (\ref{B1}) are the translates $U^a V^{-b}|\psi_{hmax}\rangle$.  
 (Remark: in the statement of Theorem 3 we have supposed that the smallest eigenvalue of $H$ is non degenerate. We expect this to be the case, but have not been able to prove it. If for some values of $\theta$ the smallest eigenvalue of $H$ is degenerate, then denote by   $|\psi_{hmin, \pm,i}\rangle$ the quantum states that are both eigenstates of $H$ with eigenvalue $h_{max}$ and eigenstates of the operator $P=\sum_{j=\bi}^{ \bs} |\bf{-j}\rangle \langle \bj |$ with eigenvalues $\pm 1$, and where $i$ labels any additional degeneracy. These states and their translates are the unique states that maximise eq. (\ref{B1}).)
}

As discussed above when $\Delta U^2$ and $\Delta V^2$ are both small, and when $d$ is large, the uncertainty relation for $U$ and $V$ reduces to the uncertainty relation for $x$ and $p$. In this limit the Hamiltonian eq. (\ref{Hamilt}) reduces to
$$
H= -(\cos \theta   + \sin \theta) I + \frac{1}{2}( \cos \theta u^2  + \sin \theta v^2)$$
and the smallest eigenvalue of $H$ is given by the smallest eigenvalue of $   \cos \theta u^2  + \sin \theta v^2$.
This suggests that we should interpret the ground states of $H$  as discrete analogues of coherent states(for $\theta = \pi/4$) and squeezed states (for $\theta \neq \pi /4$). It is this correspondence which suggests that the largest eigenvalue of $H$ is non degenerate, since the smallest eigenvalue of $ \cos \theta u^2  + \sin \theta v^2$ is non degenerate.
(This also shows that we can interpret the other eigenstates of $H$ when $\theta = \pi/4$ as discrete analogues of the number states, i.e. the eigenstates of the harmonic oscillator). Note also that in the continuous limit the operator $P$ tends to the parity operator that takes $x \to -x$ and $p \to -p$. This interpretation is discussed in detail in \cite{OBBD95,O95,OWB96,BCHKO}. We refer in particular to \cite{BCHKO} for plots of the eigenstates of $H$ when $\theta=\pi/4$ and for a discussion of how they tend to the Hermite-Gauss functions in the $d\to \infty$ limit. 
Note that the equation $H|\psi\rangle = E|\psi\rangle$  is a finite dimensional version of Harper's equation\cite{Harper}, a well studied equation in mathematical physics.

 \section{Conclusion}
 
In summary we have obtained an uncertainty relation for unitary operators $U$ and $V$ which obey the commutation relation eq. (\ref{UVVU2THEO}), which has applications to signal processing, modular variables, and the DFT. In particular in the later context this uncertainty relation generalises to the finite dimensional case the uncertainty relation for position and momentum eq. (\ref{DxDp}), and reduces to the uncertainty relation for Pauli operators eq. (\ref{DxDz}). We expect that our result will yield insights into other applications of uncertainty relations, such as the precision with which two non commuting observables can be jointly observed, or the degree to which a ''fuzzy" measurement of one observable perturbs the other observable. 

 {Acknowledgements.} We thank Igor Shparlinski and Sergei Konyagin for their crucial help  during early stages of this project.
 We acknowledge financial support by EU project QAP contract 015848, by the IAP project -Belgium Science Policy- P6/10 Photonics@be, and by the F.R.S.-F.N.R.S. under grant 2.4.548.02F.

 \appendix

 \section{The large $d$ limit and the $u$, $v$ commutator}
We use the same notation as in the main text. 

We recall that we work on a finite $d$ dimensionnal Hermitian space and we denote   $\{\vert {\bf j}\rangle\ \vert\ j=\bi,\dots,\bs\}\equiv\{\vert {\bf \bi}\rangle,\dots,\vert {\bf \bs}\rangle\}$, the basis in which the operator 
$$U=\sum_{j=\bi}^{\bs}e^{2i\pi j/d}\vert {\bf j}\rangle\langle {\bf j}\vert\qquad .$$
is diagonal.
We also  introduce a Hermitian operator $u$ such that 
$$U=e^{i u \sqrt{2\pi/d}}\qquad .$$
The action of $u$ on the basis is defined modulo a set of $d$ integers $k_j$:
$$u\ \vert {\bf j} \rangle=\sqrt{\frac {2\pi}d}(j+k_j\,d)\vert {\bf j}\rangle\equiv \nu_j\vert {\bf j}\rangle \qquad ,\qquad k_j\in \ZZ\qquad .$$
We fix a particular choice of $u$ by choosing the $k_j$ all equal to 0, 
so that the eigenvalues of $u$ belong to the interval (centered around zero) :
\begin{equation}
\nu_j\in  \left[\bi\sqrt{\frac {2\pi}d},\dots,\bs\sqrt{\frac {2\pi}d}\,\right]\qquad .
\label{nuj}\end{equation}

Let us split the set of indices into  two disjoint subsets  : 
\begin{equation}
I_{0,\delta}=\{j\vert \ \vert j \vert \leq \frac 2 \pi\left[\frac d2\right] \delta\}\quad\textrm{and}\quad J_{0,\delta}=\{j\vert \ \vert j \vert > \frac 2 \pi\left[\frac d2\right] \delta  \}\qquad \textrm{with}\qquad \delta\leq \frac \pi 2\label{IJ}
\end{equation}

On $I_{0,\delta}$ (resp. $ J_{0,\delta}$) the $u$-eigenvalues obey the inequality:
$ \vert\nu_j\vert\leq \sqrt{{2\,d} /\pi}\,\delta\ $ (resp. $ \vert\nu_j\vert> \sqrt{ {2\,d}/ \pi}\,\delta )$.
Let us define the projector
$$P_\delta=\sum_{j \in I_{0,\delta}}\vert{\bf j}\rangle\langle{\bf j}\vert\ .$$

The subsets ${\cal U}_\delta({\epsilon})$ (with, by definition, $\epsilon >0$)  are defined as the sets of vectors such that
$$\langle\psi\vert P_\delta\vert \psi\rangle\geq 1-\epsilon$$
Obviously they are $U$ invariant; indeed : $P_\delta=U^\dagger P_\delta\,U$.
Upon writing
$\vert \psi\rangle=\sum_{j}c_j\vert{\bf j}\rangle=\sum_{j\in I_{0,\delta}}c_j\vert{\bf j}\rangle +\sum_{j\in  J_{0,\delta}}c_j\vert{\bf j}\rangle$, we deduce that
\begin{equation}
(1-\epsilon)\leq \sum_{j\in I_{0,\delta}} \vert c_j\vert^2\leq 1\quad \textrm{and} \quad  \sum_{j\in J_{0,\delta}}\vert c_j\vert^2< \epsilon\ .
\label{ep}
\end{equation}

Note that we obviously have:
\begin{equation}
V^n\ {\cal U}_\delta(\epsilon)\subset {\cal U}_{\delta + \frac \pi d \,n}(\epsilon),
\label{Vn}\end{equation}
that is under a small $V$ translation ($n$ small), the parameters $\delta$ increases a little, while $\epsilon$ is not increasing.

We expect that the fact that a state belongs to ${\cal U}_\delta({\epsilon})$ and the fact that $\Delta U^2$ is small should be essentially equivalent, since both approaches measure how much the state is localised in the basis which diagonalises $U$. The main difference is that belonging to  ${\cal U}_\delta({\epsilon})$ implies that the state is centered around the eigenvalue $\nu_j=0$, whereas $\Delta U^2$ is small implies that the state is localised in the basis which diagonalises $U$, but does not say around which eigenvalue the state is centered. \\

In detail we have the following:\\

\noindent {\em Lemma 1. If $|\psi\rangle=\sum_jc_j|{\bf j}\rangle$   belongs to ${\cal U}_\delta({\epsilon})$, then $\Delta U^2\leq \frac{\delta^2}{2}+2\epsilon$ .}
\par\noindent Proof\footnote{End of proof are indicated by the symbol ${\blacksquare}$.}\ :
$$\langle \psi |U|\psi\rangle = \sum_j\cos(2\,\pi j/d)\vert c_j\vert^2 + i \sum_j\sin(2\,\pi j/d)\vert c_j\vert^2$$
Hence
$$|\langle \psi |U|\psi\rangle|^2\geq  (1-\epsilon)^2\cos^2(\delta)\geq \cos^2\delta - 2 \epsilon$$
which implies 
$$\Delta U^2 \leq 1 - (\cos^2\delta - 2 \epsilon)= 2 \sin^2\frac{\delta}{2} +  2 \epsilon \leq \frac{\delta^2}{2}+2\epsilon\qquad .\qquad\qquad {\blacksquare}$$

\noindent{\em Lemma 2. For any   $\delta >0$, if $\Delta U^2+\pi^2/d^2$ is small enough,  there exists an $\epsilon<(\Delta U^2+\pi^2/d^2)/\sin^2(\delta/2)$ and a translation $|\psi\rangle\to V^k |\psi\rangle$ such that $ V^k |\psi\rangle$  belongs to ${\cal U}_\delta({\epsilon})$}

\noindent Proof. Let us choose the translation operator $V^k$ operator, such that,   in absolute value, the phase $\alpha $ of the expectation value of 
\begin{equation}
\langle\,\psi\,\vert U\vert\,\psi\rangle=Q\, \exp[i\,\alpha] \label{Ch1}
\end{equation}
is less than $\pi/d$. Then, noticing that $Q \leq 1$, we obtain  :
\begin{eqnarray}\Delta U^2 &=& 1 - (\sum_j |c_j|^2 \cos \frac{2 \pi}{d}j)^2 - Q^2 \sin^2 \alpha\nonumber \\
&\geq& 1 - (\sum_{j\in I_{0,\delta}} |c_j|^2 \cos  \frac{2 \pi}{d} j + \sum_{j\in J_{0,\delta}} |c_j|^2 \cos \frac{2 \pi}{d}j)^2 - \frac{\pi^2}{d^2} \nonumber\\
&\geq& 1 - ( \sum_{j\in I_{0,\delta}} |c_j|^2 + \sum_{j\in J_{0,\delta}} |c_j|^2  \cos 2 \delta )^2- \frac{\pi^2}{d^2} 
\nonumber\\
&=&1 - ( 1 - \sum_{j\in J_{0,\delta}} |c_j|^2 2 \sin^2\frac{\delta}{2} )^2 - \frac{\pi^2}{d^2} \label{DUUU}
\end{eqnarray}
Let us now set 
\begin{equation}
\sin^2\beta=\Delta U^2+\frac {\pi^2}{d^2}
\end{equation}
from the last inequality (\ref{DUUU}) we deduce that
\begin{equation}
\textrm{either}\qquad\epsilon\geq\frac {\cos^2\beta/2}{\sin^2\delta/2}\qquad \textrm{or}\qquad \epsilon\leq\frac {\sin^2\beta/2}{\sin^2\delta/2}\qquad ,
\end{equation}
where we recall that $\epsilon=\sum_{j\in J_{0,\delta}}|c_j|^2$, see eq.[\ref{ep}].
Thus, if $\Delta U^2+\pi^2/d^2$ is small enough ({\it i.e.} for small $\Delta U^2$ and large $d$), since $\epsilon\leq 1$, it is the second inequality that is satified.\hfill{$ {\blacksquare} $}\strut\\

Let us now show that if a state belongs to ${\cal U}_\delta({\epsilon})$, then we can expand the operator $U$ is series, since the state lies almost entirely in the space where the eigenvalues of $u$ are small. More precisely we have :\\

\noindent{\em Lemma 3. If $|\psi\rangle$ belongs to ${\cal U}_\delta({\epsilon})$, then we can expand $U\simeq (1+i\sqrt{\frac {2\,\pi}d} u-\frac \pi d u^2)$ since
$$\left\vert U\vert\psi\rangle-(1+i\sqrt{\frac {2\,\pi}d} u-\frac \pi d u^2)\vert\psi\rangle\right\vert^2 =
O(\delta^2) +O(\epsilon )\ .$$
}

Proof.
\begin{eqnarray*}
&&\left\vert U\vert\psi\rangle-(1+i\sqrt{\frac {2\,\pi}d} u-\frac \pi d u^2)\vert\psi\rangle\right\vert^2 \\
&&=2-\langle\psi\vert U^\dagger + U\vert\psi\rangle +i\sqrt{\frac {2\,\pi}{d}}\langle\psi\vert u\,U- U^\dagger u\vert\psi\rangle  
+\frac \pi d \langle\psi\vert u^2\,U+ U^\dagger u^2\vert\psi\rangle+\frac{\pi^2}{d^2}\langle\psi\vert u^4\vert\psi\rangle
\end{eqnarray*}
But
\begin{eqnarray*}
2-\langle\psi\vert U^\dagger + U\vert\psi\rangle&=&2-2\sum_j\cos(2\,\pi j/d)\,\vert c_j\vert^2=4\sum_j\sin^2(\pi j/d)\,\vert c_j\vert^2\\
&\leq&4 \left(\sin^2(\delta)+\epsilon\right)<4\left(\delta^2 +\epsilon\right)\\
\frac \pi d \langle\psi\vert  u^2U+U^\dagger u^2\vert\psi\rangle&=&\frac {2\,\pi}d\sum_j \cos(2\,\pi\, j/d)\,\nu^2_j\,\vert c_j\vert^2\\
&\leq&\frac {2\,\pi}d \left( \frac {2\,d}{\pi}\,\delta^2+\cos(2\,\delta)\,\frac{\pi\,d}2\,\epsilon\right)<4\,\delta^2+\pi^2\,\epsilon \\
\frac{\pi^2}{d^2}\langle\psi\vert u^4\vert\psi\rangle&\leq&\frac{\pi^2}{d^2}\left( \frac{4\,d^2}{\pi^2}\,\delta^4+\frac{\pi^2\,d^2}{4}\,\epsilon\right)=4\,\delta^4+\frac{\pi^4}{4}\epsilon\\
i\sqrt{\frac {2\,\pi}{d}}\langle\psi\vert u\,U- U^\dagger u\vert\psi\rangle&=&-\sqrt{\frac {2\,\pi}{d}}\,2\sum_j\sin(2\,\pi j/d)\,\nu_j \vert c_j \vert^2\\
&=&-4\,\frac {2\,\pi}{d} \sum_j^{[\frac d2]} j\sin(2\,\pi j/d)\,  \vert c_j\vert^2\leq 0\end{eqnarray*}
Thus, collecting all these inequalities, we obtain:
$$\left\vert U\vert\psi\rangle-(1+i\sqrt{\frac {2\,\pi}d} u-\frac \pi d u^2)\vert\psi\rangle\right\vert^2\leq 4\,\delta^2+ 4\,\delta^4+(4+\pi^2 +\frac {\pi^4} 4)\epsilon \qquad .\qquad \qquad {\blacksquare}$$

Lemma 3 implies that if $|\psi\rangle$ belongs to ${\cal U}_\delta({\epsilon})$, then
we have
\begin{equation} \Delta U^2 \simeq \frac{2 \pi}{d}  (\langle u^2\rangle - \langle u \rangle^2 )\ .
\label{DUdu}\end{equation}

We now introduce the dual basis  $\{\vert \,\widetilde{\bf j}\,\rangle\ \vert\ j=-[d/2],\dots,[(d-1)/2]\}$ which diagonalises the $V$ operator:
$$V=\sum_{j=-[\frac d2]}^{\frac{d-1}2}e^{2i\pi j/d}\vert\, {\widetilde{\bf j}\,\rangle}\langle\,{ \widetilde {\bf j}}\,\vert \ .$$
We introduce the operator $v$ as 
$$V=e^{i v \sqrt{2\pi/d}}$$
where $v$ is defined by
$$v\ \vert \,{ \widetilde {\bf j}}\, \rangle = {\tilde\nu_j}\,\vert \,{ \widetilde {\bf j}}\,\rangle\qquad
\textrm{where}\qquad
{\tilde{\nu}_j}=\sqrt{\frac{2\,\pi}d}j $$
and $\nu_j$ belong to the interval eq. (\ref{nuj}).

As above, we can
use two subsets of indices [\ref{IJ}] to define the projector
$$\tilde P_\delta=\sum_{j \in  I_{0,\delta}}\vert{\widetilde {\bf  j}}\rangle\langle{\widetilde{\bf  j}}\vert\ .$$
The subsets ${\cal V}_\delta({\epsilon})$ are defined as the sets of vectors such that
$$\langle\psi\vert \tilde P_\delta\vert \psi\rangle\geq 1-\epsilon\qquad .$$

Now let us now consider the intersection of ${\cal U}_\delta({\epsilon})$ and 
${\cal V}_\delta({\epsilon})$ and show that for states belonging to this intersection, we can approximate the commutator of $u$ and $v$ by $[u,v]\simeq i$.\\

\noindent{\em Lemma 3. If $|\psi\rangle$ belongs to the intersection of ${\cal U}_\delta({\epsilon})$ and 
${\cal V}_\delta({\epsilon})$, then for large $d$ and small $\delta$, $\epsilon$, $[u,v]|\psi\rangle \simeq i |\psi\rangle$}

\noindent Proof. Let us use the identity $U^\dagger V^\dagger U V = e^{i \pi /d}\simeq 1+ i \pi/d$.
Since $|\psi\rangle$ belongs to the intersection of ${\cal U}_\delta({\epsilon})$ and 
${\cal V}_\delta({\epsilon})$, $V^\dagger U V|\psi\rangle$ belongs to ${\cal U}_{\delta + 2 \frac{\pi}{d} }({\epsilon})$, 
$U V|\psi\rangle$ belongs to ${\cal V}_{\delta + \frac{\pi}{d}  }({\epsilon})$, and 
$V|\psi\rangle$ belongs to ${\cal U}_{\delta +  \frac{\pi}{d}  }({\epsilon})$, see eq. [\ref{Vn}]. 
Hence, using Lemma 2, we can expand all the operators $U^\dagger$, $V^\dagger$, $U$, $V$ in series to obtain
$U^\dagger V^\dagger U V |\psi\rangle \simeq ( 1 + [u,v]\frac{\pi}{d} )|\psi\rangle$.\hfill{$ {\blacksquare} $}\\

The Heisenberg uncertainty principle then implies that for large $d$ and small $\delta$, $\epsilon$, states $|\psi\rangle$ belonging to the intersection of ${\cal U}_\delta({\epsilon})$ and ${\cal V}_\delta({\epsilon})$ obey
$$\Delta u \Delta v \geq \frac{1}{2}\ .$$
For these states we also have $\Delta U^2 = \frac{2 \pi}{d} \Delta u^2$
and  $\Delta V^2 = \frac{2 \pi}{d} \Delta v^2$, see eq. (\ref{DUdu}).
Thus for states belonging to the intersection of ${\cal U}_\delta({\epsilon})$ and ${\cal V}_\delta({\epsilon})$, $\Delta U^2$ and $\Delta V^2$ cannot both be arbitrarily small since they must obey
$$\Delta U^2 \Delta V^2 \geq \frac{\pi^2}{d^2}\ .
$$

It is interesting to note that we can also study the $u$, $v$ commutation relation  from another perspective. From the above definitions of the $u$ and $v$ operators, we can derive that
 \begin{eqnarray}
 \braj[u,v]\vert {\bf j}'\rangle&=&(j-j')\sqrt{\frac{2\,\pi}d}\braj v \vert {\bf j}'\rangle\nonumber\\
 &=&i\,(-1)^{(j-j'+1)}\frac{\pi(j-j')/d}{\sin[\pi(j-j')/d]}\times \left\{
 \begin{array}{l} 1\ \rm{ if }\  d=2\,n +1\\
 e^{-i\pi(j-j')/d}\ \rm { if }\  d=2\,n \end {array}\right.\label{juvj}
 \end{eqnarray}
  We have studied the matrix [\ref{juvj}] numerically and have observed that, for large values of $d$, a quite remarkable property holds: {\em almost all the eigenvalues are closed to $i$.}\footnote{But not all; indeed since the trace of the matrix vanishes the few other eigenvalues have to sum to minus the sum of those close to $i$.} For instance, for $d=801$, we found that the matrix $i[u,v]$ has $61\%$ of its eigenvalues in the range $1-10^{-10}$ and $1+10^{-10}$. In other words, there exists a very large subspace of the initial 801-dimensional space on which the $u$, $v$ commutator is nearly equal to $i$. We have however not been able to derive this result analytically.  \\

Finally let us discuss briefly the states that lie in the intersection of ${\cal U}_\delta(\epsilon)$ and ${\cal V}_{\delta'}(\epsilon')$. In terms of $u$ and $v$ variables, these states are localized near $u=0$ and $v=0$. Furthermore for these states, we have $[u,v]\simeq i$. 
But if we take continuous variables $x$ and $p$ obeying $[x,p]=i$, then it is easy to find states located in the vicinity of $x=0, p=0$. For instance Gaussian states. We expect that discretized versions of these continuous states located near $x=0, p=0$ should lie in the intersection of ${\cal U}_\delta(\epsilon)$ and ${\cal V}_{\delta'}(\epsilon')$. 

This intuition is indeed born out, and, for large $d$, we checked numerically that the discretized gaussian
\begin{equation}
\vert{\bf  \Gamma_\sigma}\rangle=\frac 1 {{\cal N}_\sigma}\sum e^{-\frac {2\,\pi\,\nu_j^2}{\sigma^2\,d}}\vert{\bf j}\rangle\qquad \textrm{with}\qquad {\cal N}_\sigma^2\simeq\sigma\,\sqrt{\frac d2}
\end{equation}
lies in the intersection of  ${\cal U}_\delta(\epsilon)$ and ${\cal V}_{\delta}(\epsilon)$ for suitable choices of $\sigma$.
 
\section{Proof of Theorem 1}
 {\em Theorem 1: Consider two unitary operators $U$ and $V$ which obey
\begin{equation}
UV = VU e^{i \Phi}
\quad,\quad
U^\dagger V = V U^\dagger e^{-i \Phi}
\quad,\quad 0\leq \Phi\leq \pi
\label{UVVU2THEOB}\end{equation}
and define
\begin{eqnarray}
\Delta U^2 = 1 - |\langle \psi |U |\psi \rangle |^2 
\quad,\quad
\Delta V^2 = 1 - |\langle \psi |V |\psi \rangle |^2
\end{eqnarray}
which are trivially bounded by 
$ 0 \leq \Delta U^2 \leq 1$, $0 \leq \Delta V^2 \leq 1
$
and let
\begin{equation}
A=\tan \frac{\Phi}{2} \quad  , \quad 0 \leq A \leq + \infty \ .
\end{equation}
Then we have the bound
\begin{equation}
(1 + 2 A) \Delta U^2 \Delta V^2 
+ A^2 (\Delta U^2 + \Delta V^2) \geq A^2 \ .
\label{THEOB}
\end{equation}
}
  
 \noindent Proof: To prove eq. (\ref{THEOB}) let us introduce the sine and cosine operators
 (for previous uses of such operators in the context of uncertainty relations see \cite{NC2,CarruthersNieto}):
 \begin{eqnarray}
 C_U = \frac{U + U^\dagger}{2} &\quad&
  S_U = \frac{U - U^\dagger}{2i} \nonumber\\
   C_V = \frac{V + V^\dagger}{2} &\quad&
    S_V = \frac{V - V^\dagger}{2i} \label{SC}
    \end{eqnarray}
 These operators are hermitian: and obey
 $
 C_U^2 + S_U^2 = C_V^2+S_V^2= \one
 $.
 We can rewrite
 $$
 \Delta U^2 = \Delta C_U^2 + \Delta S_U^2 \ .$$
 And then, using the Robertson inequality
 \begin{equation}
\Delta A \Delta B \geq \frac{1}{2} | \langle \psi | [A,B] | \psi \rangle| \ ,
\label{DADB-B}
\end{equation}
 which holds for all hermitian $A$ and $B$, we obtain the bound
 \begin{equation}
 \Delta U^2 \Delta V^2 \geq
 \Delta S_U^2\Delta S_V^2 \geq
\frac{1}{4} | \langle \psi | [S_U , S_V] | \psi \rangle |^2 \ .\qquad \qquad  {\blacksquare}
\label{RDUDV}
\end{equation}

We now prove the following result:\\

\noindent{\em Lemma 1: For any unitary operators obeying eqs. (\ref{UVVU2THEOB}), and sine and cosine operators given by eqs. (\ref{SC}) we have
\begin{equation}
[S_U,S_V]= -i \tan \frac{\Phi}{2} (C_U C_V + C_V C_U)\qquad .
\label{SS}
\end{equation}
}
 \\
\noindent {\em Proof:} We can write the products of sine and cosine operators as
 \begin{eqnarray}
 S_U S_V &=& -\frac{1}{4} ( U V + U^\dagger V^\dagger) 
 + \frac{1}{4} ( U^\dagger V + U V^\dagger) \label{1}\\
 &=& -\frac{e^{i \Phi} }{4} ( V U + V^\dagger U^\dagger) 
 + \frac{e^{-i \Phi}}{4} ( V^\dagger U + V U^\dagger)\quad \label{2}\\
  S_V S_U &=& -\frac{1}{4} ( V U + V^\dagger U^\dagger) 
 + \frac{1}{4} ( V^\dagger U + V U^\dagger) \label{3}\\
 &=& -\frac{e^{-i \Phi} }{4} (  U V + U^\dagger V^\dagger) 
 + \frac{e^{+i \Phi}}{4} ( U V^\dagger  + U^\dagger V) \quad\quad\label{4}\\
 C_U C_V &=& \frac{1}{4} ( U V + U^\dagger V^\dagger) 
 + \frac{1}{4} ( U^\dagger V + U V^\dagger) \label{5}\\
  C_V C_U &=& \frac{1}{4} ( V U + V^\dagger U^\dagger) 
 + \frac{1}{4} ( V^\dagger U + V U^\dagger) \label{6}
 \end{eqnarray}
 Taking the difference of eqs. (\ref{2}) and (\ref{3}), and then using eqs. (\ref{3}) and (\ref{6}) we obtain
 \begin{eqnarray}
 [S_U,S_V] &=&
 -\frac{1}{4} (e^{i \Phi} -1) ( V U + V^\dagger U^\dagger) +\nonumber\\
 & & \quad
 \frac{1}{4} (e^{-i \Phi} -1)( V^\dagger U + V U^\dagger)\nonumber\\
 &=& - i \sin \Phi C_V C_U - 2 \sin^2 \frac{\Phi}{2} S_V S_U\nonumber
 \end{eqnarray}
 Similarly using eqs. (\ref{1}) and (\ref{4}) we get
 \begin{eqnarray}
 [S_U,S_V] 
 &=& - i \sin \Phi C_U C_V + 2 \sin^2 \frac{\Phi}{2} S_U S_V\nonumber
 \end{eqnarray}
 Combining these two expressions yields 
 $$
 [S_U,S_V] = - i \frac{\sin \Phi}{2 \cos^2 \frac{\Phi}{2}} (C_U C_V + C_V C_U)\qquad .\qquad \qquad  {\blacksquare} $$ 
  
 To proceed recall that if we change the phase of $U$ and $V$: $U\to e^{i\mu}U$, $V\to e^{i\mu}V$, then the uncertainties $\Delta U$ and $\Delta V$ do not change. Let us choose these phases so that $\langle \psi | U | \psi \rangle$ and  $\langle \psi | V | \psi \rangle$ are real and positive. We then have $\langle \psi | S_U | \psi \rangle = \langle \psi | S_V | \psi \rangle=0$, and we can prove :
 \\ \strut\\
\noindent {\em Lemma 2: With a choice of phase for the operators $U$ and $V$ such that 
 $\langle \psi | C_U | \psi \rangle$, $\langle \psi | C_V | \psi \rangle$  are real and positive, which implies $\langle \psi | S_U | \psi \rangle = \langle \psi | S_V | \psi \rangle=0$, 
 we have
 \begin{equation}
 |\langle \psi | C_U C_V | \psi \rangle | \geq \sqrt{1 - \Delta U^2}\sqrt{1 - \Delta V^2}
 - \Delta U \Delta V
 \label{L2}
 \end{equation}
 }
 \\
 \noindent{\em Proof:} With the above choice of phase we have
 $$\Delta U^2 = 1 - \langle \psi | C_U | \psi \rangle^2\qquad .$$
 We can also write
 $C_U |\psi\rangle = x_U |\psi\rangle + y_U |\psi^\perp\rangle
$
where $|\psi^\perp\rangle$ is a normalised quantum state orthogonal to $|\psi\rangle$,
$x_U= \langle \psi | C_U | \psi \rangle$, and $x_U^2 + y_U^2 \leq 1$ (since all the eigenvalues of $C_U^2$ are less or equal to $1$). Hence $y_U^2 \leq \Delta U^2$.
Similarly we have
$\Delta V^2 = 1 - \langle \psi | C_V | \psi \rangle^2$
 and
 $C_V |\psi\rangle = x_V |\psi\rangle + y_V |\psi'^\perp\rangle
$
where $|\psi'^\perp\rangle$ is a quantum state orthogonal to $|\psi\rangle$,
$x_V= \langle \psi | C_V | \psi \rangle$, $x_V^2 + y_V^2 \leq 1$, and $y_V^2 \leq \Delta V^2$.
 
 \noindent Putting these expressions together we have :
 \begin{eqnarray}
 |\langle C_V C_U \rangle | &=& |x_U x_V + y_U y_V \langle \psi'^\perp |\psi \rangle|\nonumber\\ 
 &\geq& |x_U x_V| - |y_U y_V| | \langle \psi'^\perp |\psi \rangle|\nonumber\\ 
 &\geq& \sqrt{1 - \Delta U^2}\sqrt{1 - \Delta V^2}
 - \Delta U \Delta V \ .
 \end{eqnarray}
{\hfill $ {\blacksquare}$}
 
 To prove our main result we insert eq. [\ref{L2}] into eq. [\ref{SS}], and the resulting expression into eq. [\ref{RDUDV}] to obtain
 \begin{equation}
 \Delta U  \Delta V \geq \tan \frac{\Phi}{2} ( \sqrt{1 - \Delta U^2}\sqrt{1 - \Delta V^2}
 - \Delta U \Delta V )\qquad .
 \end{equation}
 Reorganising terms and squaring yields eq. (\ref{THEOB}). \hfill$ {\blacksquare}$

\section{Proof of Theorems 2 and 3}

To prove theorems 2 and 3, note that
$$ \langle \psi | U | \psi \rangle  =  \sum_j |c_\bj |^2 e^{+i 2 \pi j  /d} $$
and
$$ \langle \psi | V | \psi \rangle  =  \sum_j \overline c_{j+1} c_j  $$
Hence if we change the phase of $c_j$: $c_j \to e^{i \varphi_j} c_j$, then $ \langle  U  \rangle $ remains unchaged, but $ \langle  V  \rangle $ changes.
The maximum value of $| \langle  V  \rangle |$ with $|c_\bj |$ fixed occurs when the phase of $\overline c_{j+1} c_j $ is independent of $j$. This implies that 
$$
c_j = e^{i \varphi_0} e^{+i 2 \pi j a /d} |c_\bj |$$
for some $a=0,...,d-1$. Hence the maximum  of $| \langle  V  \rangle |$ at fixed 
$ \langle  U  \rangle $ occurs when $ \langle  V  \rangle = | \langle  V  \rangle | e^{-i 2 \pi  a /d}$.

We next repeat this argument, but working in the dual basis $\tilde c_k$. In this way we obtain that the maximum of 
$| \langle  U  \rangle |$ at fixed 
$ \langle  V  \rangle $ occurs when $ \langle  U  \rangle = | \langle  U  \rangle | e^{-i 2 \pi  b /d}$ for some $b=0,...,d-1$.

Applying the above two arguments successively, we deduce that the maximum of 
\begin{equation}
\cos \theta |\langle \psi| U  |\psi \rangle| + \sin \theta |\langle \psi |  V |\psi  \rangle| , \quad 0\leq \theta \leq \pi/2
 \label{B1B}
\end{equation}
occurs on states such that 
$ \langle  U  \rangle = | \langle  U  \rangle | e^{-i 2 \pi  b /d}$ and $ \langle  V  \rangle  = | \langle  V  \rangle | e^{-i 2 \pi  a /d}$. Applying a translation $U^aV^{-b}$, does not change $| \langle  U  \rangle |$ and $| \langle  V  \rangle |$. Hence we can search for the maximum of eq. (\ref{B1B}) among the states with $ \langle  U  \rangle $ and $ \langle  V  \rangle $ both real positive.

For states with this property, we have
$ \langle  U  \rangle =  \langle  C_U  \rangle $
and
$ \langle  V  \rangle =  \langle  C_V  \rangle $
where $C_U= (U + U^\dagger)/2$ and $C_V= (V + V^\dagger)/2$.

Hence we can replace the maximisation of eq. (\ref{B1B}) by:
\begin{equation}
\mbox{maximise } \cos \theta \langle \psi| C_U |\psi \rangle + \sin \theta \langle \psi |  C_V |\psi  \rangle 
 \label{B3}
 \end{equation}
  under the conditions $ \langle  U  \rangle $ and $ \langle  V  \rangle $  both real.

Let us drop for the moment the last two conditions. Then the maximum will be given by the largest eigenvalue of the Hermitian operator 
 \begin{equation}
 H= -\cos \theta C_U  - \sin \theta C_V  , \quad 0\leq \theta \leq \pi/2
 \label{HamiltB}
 \end{equation}
 We denote this largest eigenvalue by $h_{max}$.

Let us now consider the hermitian operator
\begin{equation}
P=\sum_{j=\bi}^{ \bs} |\bf{-j}\rangle \langle \bj |\label{Pop}
\end{equation}
Note that $P^2=I$ which implies that the eigenvalues of $P$ are $\pm 1$, and that
 we have
 \begin{eqnarray}
 PUP&=&U^\dagger\label{PU}\\
 PVP&=&V^\dagger \label{PV}
 \end{eqnarray}
 Hence if $|\psi\rangle$ is a an eigenstate of $P$ then 
  $\langle \psi| U  |\psi \rangle$ and $\langle \psi |  V |\psi  \rangle$ are both real.
Furthermore 
 $P$ commutes with $H$, and we can diagonalise simulataneously $H$ and $P$.
 Let us denote by   $|\psi_{hmax,p}\rangle$ , with $p=\pm 1$ the quantum states that are both eigenstates of $H$ with eigenvalue $h_{max}$ and eigenstates of the operator P (eq.[\ref{Pop}]) with eigenvalue $p$. These are the unique states 
 which maximise eq. (\ref{B1B}) and have $ \langle  U  \rangle $ and $ \langle  V  \rangle $  both real. The only other states which maximise eq. (\ref{B1B}) are the translates of $|\psi_{hmax,i}\rangle$. This concludes the proofs of theorems 2 and 3.
 
As a final remark, note also that one can always choose the joint eigenstates of $H$ and $P$ to be real, i.e. choose $c_j= \langle \bj |\psi_{h_{max},i}\rangle$ to be real. Indeed if $|\psi\rangle$ is an eigenstate of $H$ and $P$, then $|\overline \psi\rangle$ is also, hence the real and imaginary parts of $|\psi\rangle$ are also. When the eigenvalue of $H$ is non degenerate, one can also take the $\tilde c_k$ real.


\end{document}